\documentstyle[preprint,aps,epsf]{revtex}             
\input epsf
\begin{document}
\pagestyle{empty}                                      
\preprint{
\font\fortssbx=cmssbx10 scaled \magstep2
\hbox to \hsize{
\hfill$\raise .5cm\vtop{
\hbox{Alberta-Thy-08-97
}\hbox{April, 1997}}$}
}
\draft
\vfill
\title{ Final-State Interactions in Heavy Quark Decays}

\author{A.N.Kamal and C.W.Luo}
\address{Theoretical Physics Institute and Department of Physics,\\ University of Alberta,
Edmonton, Alberta T6G 2J1, Canada
}

%
%
\vfill
\maketitle
\begin{abstract}
We study the role played by the final-state interactions (fsi) in heavy quark decays using Pomeron and Regge exchanges to describe high energy scattering.  At center of mass energy $\sqrt{s}\sim 5GeV $,  Pomeron dominance does not apply. We study the behavior of the decay amplitudes as $\sqrt{s} $ is varied close to the B mass. We also investigate the behavior
of the decay amplitude as $\sqrt{s}\rightarrow \infty $.  Our conclusion is that the decay amplitudes approach a real value asymptotically.
\end{abstract}
%
%

%
%
\pagestyle{plain}

\par
Recently,  much has been written\cite{Zheng,DGPS,NP,BH,BGR} about  the role played by the final state interactions (fsi) in heavy quark decays. The conclusions are contradictory: Ref \cite{DGPS} claims that fsi continue to play an important role in B decays by imparting non-vanishing phases to the decay amplitudes ( a conclusion the authors of \cite{BH,BGR} exploit ) while Refs\cite{Zheng} and \cite{NP} arrive at  an opposite view that the phases generated by fsi are insignificantly small.

\par

In the present work, we have analyzed the question anew to achieve a better understanding of the role of fsi at B mass and higher. We begin with a few relevant details  which also assist in establishing   the notation. The first point to realize is that at B mass, center of mass energy $\sqrt{s}=m_B $, the total cross-sections for the most hadronic reactions ($K^+p, K^+n $ excepted ) are decreasing with energy\cite{RMB}. Thus Pomeron dominance does not set in at $\sqrt{s}\approx 5GeV $. Regge exchanges play  an important role at B mass, though there is an energy at which Pomeron dominance sets in signaling  the rise in total cross-sections.  This occurs at  energies  above  $\sim15GeV $.

\par
In order to follow the reasoning of  \cite{DGPS}, define the partial-wave decomposition of scattering amplitude, T(s, t), for dissimilar spin-less particles as follows,
\begin{equation}
T(s, t)=\frac{8\pi\sqrt{s}}{k}\sum{(2l+1) A_l(s) P_l(cos\theta}),
\end{equation}
where  k is the center of mass momentum and $A_l(s) $ the partial-wave amplitudes. The discontinuity  for the two-body decay amplitude, $M(s) $, for $ B\rightarrow f $, where f involves spin-less particles, arising solely from the elastic channel is \cite{DGPS}
\begin{equation}
\sigma (s) \equiv Disc M(s)=\frac{1}{2}\frac{1}{(2\pi)^2}\int \frac{d^3\vec{p}}{(2E_1)(2E_2)}\delta(\sqrt{s}-E_1-E_2) T^*(s,t) M(s),
\end{equation}
where $T(s, t) $ is the amplitude for $f\rightarrow f$ scattering. Using (1)  in (2)  one gets
\begin{equation}
\sigma(s)= A^*_0(s) M(s).
\end{equation}
 In what follows we ignore the spin and isospin complications as they are inessential to our arguments. Because of this  only $A_0(s) $ will appear in our formulation.  If M(s) is a real analytic function satisfying a dispersion relation, the spectral function $\sigma(s) $ is real. For a single-channel problem, the S-wave amplitude $A_0(s) $ is  $ e^{i\delta_0}sin\delta_0 $.  Thus the decay amplitude in (3) must have a phase $\delta_0$ in order to generate a real $\sigma(s) $.  This is the content of Watson's theorem\cite{Watson}. In general, when other channels are present, $A_0(s)=(\eta e^{2i\delta_1}-1)/2i $, where  $\delta_1 $ is an eigenphase and the elasticity is limited by $0\leq \eta \leq 1 $. Reality of $\sigma(s) $ now requires additional contributions to the right hand side of   (3) which we discuss in the following.

\par
To be specific we consider the decay $B\rightarrow D^*\pi $ but  ignore the spin and isospin of the final state particles only to the extent needed. As noted in \cite{NP}, $D\rho $ channel will mix with $D^*\pi $,  but   $D^*\pi^{\prime} $ and $D\rho^{\prime} $ channels,  where $\pi^{\prime} $ and $\rho^{\prime} $ are higher mass  particles with the same quantum numbers as $\pi $ and $\rho $ respectively,  could also mix with $D^*\pi $ state. Indeed,   intermediate  states $D^*(n\pi) $  enter the picture, where for n odd one could form a $\pi $-like  and for n even a $\rho$-like combination.  A generalization of   (3) for the spectral function is \begin{eqnarray}
\sigma(s)\equiv & Disc M_{B\rightarrow D^*\pi}(s) = A^*_0(D^*\pi\rightarrow D^*\pi) M_{B\rightarrow D^*\pi}(s) + A^*_0(D^*\rho\rightarrow D^*\pi) M_{B\rightarrow D^*\rho}(s)+ &  \nonumber \\
 &A^*_0(D^*\pi^{\prime}\rightarrow D^*\pi) M_{B\rightarrow D^*\pi^{\prime}}(s)+A^*_0(D\rho^{\prime}\rightarrow D^*\pi) M_{B\rightarrow D\rho^{\prime}}(s) 
+.......  &
\end{eqnarray}

\par
Let us now  assume that a Regge-exchange  description applies to the scattering amplitudes.  (Ref \cite{Zheng} has concluded that cuts do not play an important role. )  The elastic channel $D^*\pi\rightarrow D^*\pi $ receives contributions from the Pomeron  (P), and the $\rho-$ and $f-$ Regge trajectories. The inelastic scatterings $D\rho\rightarrow D^*\pi $ and $D\rho^{\prime}\rightarrow D^*\pi $ receive contributions only from $ \pi $ or $ \pi^{\prime} $ Regge trajectories. However,  the inelastic scattering  $D^*\pi^{\prime}\rightarrow D^*\pi $,  as well  as $D\rho\rightarrow D\rho^{\prime} $,  receives contributions from P,  f,  $\rho $ and $\rho^{\prime} $  trajectories.   Let us now study the S-wave amplitudes and their behavior  as a function of s.

\par
The Pomeron contribution to the elastic channel, $D^*\pi\rightarrow D^*\pi $ , or the inelastic channel   $D^*\pi^{\prime}\rightarrow D^*\pi $, can be parameterized in a form (we assume that Pomeron amplitude is absorptive)
\begin{equation}
T^P(s, t)= i\beta (\frac{s}{s_0})^{1.08} e^{b_P t},  
\end{equation}
where the Pomeron trajectory is represented \cite{DGPS,DL} by $\alpha_P(t)=1.08+\alpha^{\prime}_P t $, $b_P=\alpha^{\prime}_Pln(\frac{s}{s_0}),  \alpha^{\prime}_P=0.25GeV^{-2} $, and $t\approx -s (1-cos\theta )/2 $ ( s and t are expressed in $ GeV^2 $ with $ s_0=1   GeV^2 $).  Projecting the S-wave amplitude from (5)  in the limit where all masses can be ignored, one obtains
\begin{equation}
A^P_0(s)\approx \frac{i\beta}{16\pi b_P s_0}(\frac{s}{s_0})^{0.08}.
\end{equation}
Typically, for $\beta\approx 10 $\cite{Zheng,NP},  $A^P_0(s)\approx 0.3 i $ at $\sqrt{s}= 5 GeV $.  The important point is that  $A^P_0(s) $ is purely imaginary and small compared to unity. Because of the $ s $ dependence of $b_P $ in  (6),  $A^P_0(s) $ decreases slowly with $ s $  along  the imaginary axis of the Argand diagram, Fig.1,  reaching a minimum, of approximately half of its value at $\sqrt{s}= 5 GeV $, at $\sqrt{s}\approx 500 GeV $.   $A^P_0(s) $ then increases slowly eventually violating the unitary limit, $Im A_l(s)\leq 1 $, at $\sqrt{s}\approx 2.5\times 10^{11} GeV $ for $\beta=10 $. This, of course,  implies that $\alpha_P(0) $ cannot remain at 1.08 for all energies, and must eventually decrease.   Ref. \cite{DL} makes the argument for "softening" of the Pomeron due to the  violation  of the Froissart bound\cite{MF} on the total cross-section albeit at an energy several orders  of magnitude higher.   Around the B mass, say, $\sqrt{s}\sim 5~ to~ 10 GeV $, the Pomeron partial-wave amplitude, $A^P_0 $, is purely imaginary and decreases with energy.
\par
Consider now the Regge-exchange amplitudes.  As a typical example, the $\rho$-exchange amplitude\cite{BH} is represented as,
\begin{equation}
T^{(\rho )}(s, t)=\frac{\beta(0)}{\Gamma (\alpha_{\rho}(t))}\frac{(1-e^{-i\pi\alpha_{\rho}(t)})}{sin\pi\alpha_{\rho}(t)} (\frac{s}{s_0})^{\alpha_{\rho}(t)}.
\end{equation}
For simplicity, assume $\alpha_{\rho}(t)= 0.5 +\alpha_{\rho}^{\prime} t $, with  $\alpha_{\rho}^{\prime} \approx 1 GeV^{-2} $ and \cite{BH}  $\Gamma (\alpha_{\rho}(t)) sin(\pi\alpha_{\rho}(t))\approx \Gamma (\alpha_{\rho}(0)) sin(\pi\alpha_{\rho}(0)) \approx \sqrt{\pi} $.  We then obtain the S-wave amplitude as 
\begin{equation}
A^{(\rho)}_0(s)\approx \frac{\beta(0)}{ 32\pi^{3/2}\alpha^{\prime}_{\rho} s} (\frac{s}{s_0})^{\alpha_{\rho}(0)} (0.30+ i 0.32).
\end{equation}
For $\beta(0)\approx 100 $\cite{BH} and at $ \sqrt{s}\approx 5 GeV $, one finds, first, that this amplitude is much smaller than the Pomeron-generated amplitude, the real and imaginary parts being $\approx 0.035 $  and, second, that $A_0^{(\rho)}(s) $ decreases with energy as  $ (\frac{s}{s_0})^{\alpha_{\rho}(0)-1}/ ln (\frac{s}{s_0} )  $.  Thus in the Argand diagram of Fig.1, while the Pomeron amplitude moves slowly downwards along the imaginary axis with increasing energy around the B mass,  the Regge amplitudes move towards the base ("no scattering" limit ) of the unitary circle much faster.


\let\picnaturalsize=N
\def\picsize{3.0in}
\def\picfilename{fig.eps}
\ifx\nopictures Y\else{\ifx\epsfloaded Y\else\input epsf \fi
\let\epsfloaded=Y
\centerline{\ifx\picnaturalsize N\epsfxsize \picsize\fi \epsfbox{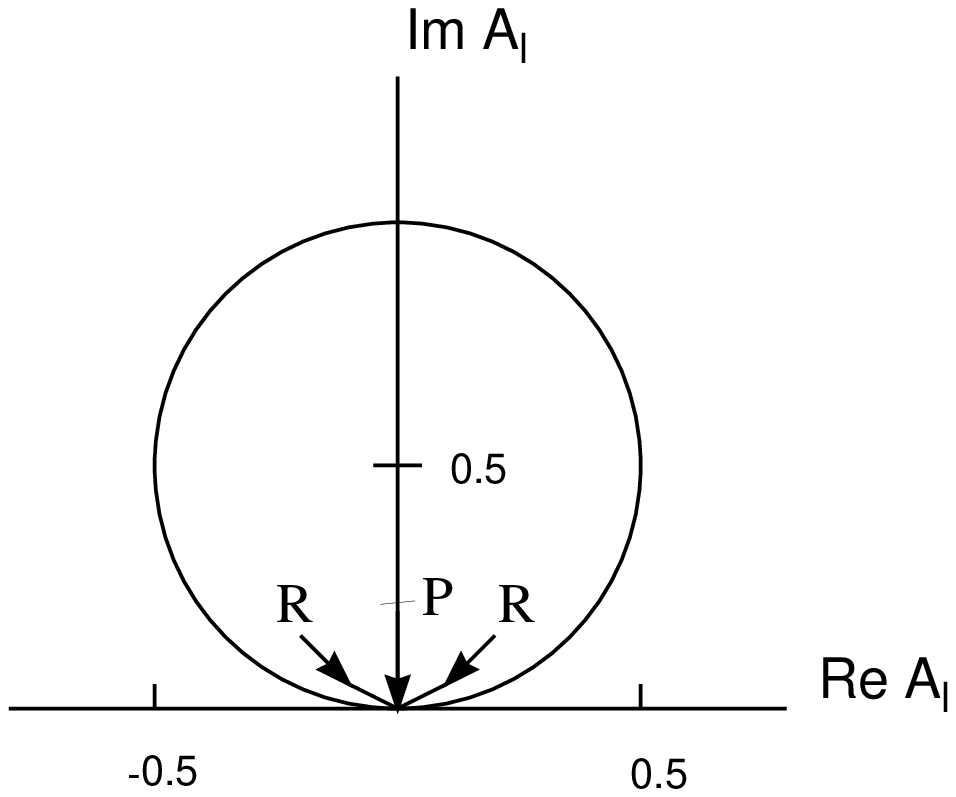}}}\fi
\begin{small}
{\hskip -0.8cm} 
Fig.1 The partial-wave amplitudes in Argand diagram.  P represents the Pomeron-generated,
and  R  the Regge-generated amplitude.  The arrows indicate the motion of the amplitudes with increasing $s $ (for $\sqrt{s} < 500 GeV $). \\
\end{small}
\par

The foregoing analysis allows us  to understand the conclusion reached in Refs\cite{Zheng} and \cite{NP}.   Define the $l=0 $  S-matrix as  ( P and R stand for Pomeron and Regge respectively),
\begin{equation}
\bf{S}_{l=0}(s)=\bf{1}+2i (\bf{A}^P_0(s)+ \bf{A}^R_0(s)).
\end{equation}

 As discussed above,  $Im A^P_0(s) $ and  $Im A^R_0(s) $ are both small but positive.  Thus, in the B mass region, the real parts of the diagonal elements of $\bf{S} $  are close to, but less than, unity. The imaginary parts of  the diagonal elements,  arising entirely from $A^R_0(s) $, are  necessarily  small.  In the schemes of \cite{Zheng} and \cite{NP}, the off-diagonal elements of the S-matrix arise entirely from  $A^R_0(s) $ and, therefore, are small.  As a consequence, the two-channel unitarity is not satisfied.    A  difference between our  formulation   and those of \cite{Zheng} and \cite{NP} is that  we include off-diagonal channels, $D^*\pi^{\prime}\rightarrow D^*\pi $  and $D\rho^{\prime}\rightarrow D\rho $  etc., which are  Pomeron driven and could,  in principle,  dominate  the  off-diagonal elements of the $S $-matrix.

\par
As we increase the energy $ but~ still~  keeping~ it~ in~ the~ B-mass~ region $, the real parts of the diagonal elements of the S-matrix rise towards unity and the imaginary parts decrease. The coupled channel problem involving channels $D^*\pi, D^*\pi^{\prime}, D^*\pi^{\prime\prime}, ...., D\rho, D\rho^{\prime}, D\rho^{\prime\prime},...,  $ tends towards a block-diagonal form with 
 $D^*\pi,  D^*\pi^{\prime}, D^*\pi^{\prime\prime},  $...., states forming  one block and $D\rho,  D\rho^{\prime}, D\rho^{\prime\prime} $, ...,  composing the other. The transitions between these blocks,  which proceed  only through Regge exchanges,  get weaker with increasing energy. The transitions  within each block  being Pomeron dominated result in largely real   S-matrix  elements (if the Pomeron is purely absorptive).
\par
The question: what happens in the limit $s\rightarrow \infty $? is harder to answer.  As we saw, this limit cannot be taken for the Pomeron amplitude with $\alpha_P(0)=1.08 $. However, if we go to an energy large yet smaller than the value where violation of partial-wave unitary sets in, 
the Regge-driven $A^R_0(s) $ approach the base of the unitary circle of Fig.1 and the Pomeron-driven amplitudes remain on the imaginary axis.  The resultant S-matrix approaches a real-symmetric form. If, further, the Pomeron "softens" to the extend that $\alpha_P(0)\rightarrow 1 $ asymptotically, then $ A^P_0(s) $ also approaches the base of the unitary circle and the S-matrix approaches the unit matrix ("no-scattering" limit).

 \par
Let us return to the case of B decays and study the energy dependence around the B mass region.

\par
First,   if for two-body channels,
\begin{equation}
\bf{M}(s)= (1-i \bf{k}^{\frac{1}{2}}\bf{K}\bf{k}^{\frac{1}{2}})^{-1} \bf{M}^{(0)}(s),
\end{equation}
where $M^0_i $ are real, $\bf{k} $ is a diagonal momentum matrix and $\bf{K} $ a real-symmetric matrix, and the unitary S-matrix is  given by \cite{RGN}
\begin{equation}
\bf{S}=(1+i\bf{k}^{\frac{1}{2}}\bf{K}\bf{k}^{\frac{1}{2}}) (1-i\bf{k}^{\frac{1}{2}}\bf{K}\bf{k}^{\frac{1}{2}})^{-1},
\end{equation}
then it is easy to show that 
\begin{equation}
\bf{M}=\frac{1}{2} (\bf{1}+\bf{S} ) \bf{M}^{(0)},
\end{equation}
for any number of two-body channels.

\par
Next we show that (12) ensures reality of the spectral function matrix  $\bf{\sigma (s) }$ of (4). 
Using (10) and (12) in (4),  we obtain
\begin{eqnarray}
\bf{\sigma}(s) & = &\frac{1}{4i}(\bf{1}-\bf{S}^+)(\bf{1}+\bf{S}) \bf{M}^0(s) \nonumber  \\
  			& = & \frac{1}{4i} (\bf{S}-\bf{S}^+) \bf{M}^0(s),
\end{eqnarray}
which is real.  As $\bf{S} $ approaches a real-symmetric form, $\bf{S}^+\rightarrow \bf{S} $, clearly the spectral function $\bf{\sigma}(s) $ approaches zero.   Eq.(12) is not the only form that ensures reality of the spectral function.   If we assume
\begin{equation}
\bf{M}(s)=\bf{S}^{\frac{1}{2}} \bf{M}^0(s),
\end{equation}
one finds
\begin{equation}
\bf{\sigma}(s)=\frac{1}{4i}(S^{\frac{1}{2}}-S^{+\frac{1}{2}})\bf{M}^0(s)
\end{equation}
which is also real. We comment later on (14).   To see how (13) works and, in particular, to derive the results of \cite{DGPS}, let us specialize first to a 2-channel problem. The S-matrix is represented in terms of 3 real parameters (in general, an n-channel S matrix is expressed in terms of $\frac{n}{2}(n+1) $ real parameters),
\begin{equation}
\bf{S}=\pmatrix{
\eta e^{2i\delta_1} & i (1-\eta^2)^{\frac{1}{2}} e^{i (\delta_1+\delta_2)} \cr
i (1-\eta^2)^{\frac{1}{2}} e^{i (\delta_1+\delta_2)} & \eta e^{2i\delta_2} \cr
}.
\end{equation}
where 
$\delta_1 $ and $\delta_2 $ are the eigenphases and  $\eta $,  the elasticity, is limited by $0\leq \eta\leq 1  $. Using (16) in (13), we obtain
\begin{equation}
\sigma_{1, 2}(s) = \frac{1}{2}\eta sin2\delta_{1, 2} M^0_{1, 2}(s)+\frac{1}{2}(1-\eta^2)^{\frac{1}{2}}cos(\delta_1+\delta_2) M^0_{2, 1}(s)  
\end{equation}
which are real.  In addition,  using (16) in (12), we get
\begin{eqnarray}
Im M_{1, 2}(s)=\sigma_{1, 2}(s) ~~~~~~~~~~~~~~~~~~~~~~~~~~~~~~~~~~                                                                                            \\
Re M_{1, 2}(s)=\frac{1}{2}(1+\eta cos2\delta_{1, 2} ) M_{1, 2}^0-\frac{1}{2}(1-\eta^2)^{\frac{1}{2}}sin(\delta_1+\delta_2) M^0_{2, 1}.          \nonumber
\end{eqnarray}
The result of \cite{DGPS} that
\begin{equation}
\frac{Im M_1(s)}{Re M_1(s)}=\sqrt{\epsilon}\frac{M^0_2(s)}{M^0_1(s)}
\end{equation}
is recovered in the limit $\delta_{1, 2}=0 $ and $\eta=1-2\epsilon $ with $\epsilon $ small and positive.  The argument of \cite{DGPS} is that as inelasticity is needed to ensure reality of the spectral function, $\epsilon $ can not vanish  and $M_1(s) $ is, therefore, complex implying that fsi effects do not disappear as the energy increases.  In the  following, we explore this issue further.

This we do in the  framework of a 3-channel problem, the reason being that apart from pedagogy, it allows for a more general form of the S-matrix---the  off-diagonal  elements of a
1-parameter representation of a 2-channel problem are purely imaginary while their 3-channel analogues are complex.

The most general form of S-matrix in a 3-channel problem can be written as,
\begin{equation}
S=\pmatrix{
\eta_1e^{2i\delta_1}  & ( \rho_{12}+i\xi_{12} )e^{i (\delta_1+\delta_2)} & ( \rho_{13}+i\xi_{13} )e^{i (\delta_1+\delta_3)} \cr
( \rho_{12}+i\xi_{12} )e^{i (\delta_1+\delta_2)} & \eta_2e^{2i\delta_2}  &  ( \rho_{23}+i\xi_{23} )e^{i (\delta_2+\delta_3)}  \cr
 ( \rho_{13}+i\xi_{13} )e^{i (\delta_1+\delta_3)} &  ( \rho_{23}+i\xi_{23} )e^{i (\delta_2+\delta_3)}  & \eta_3e^{2i\delta_3}  \cr
}
\end{equation}
where with $i\neq j\neq k $,
\begin{eqnarray}
\rho_{ij} &= &-\frac{1}{4\sqrt{\eta_i\eta_j}}\sqrt{(1+\eta_k+\eta_i-\eta_j) (1+\eta_k-\eta_i+\eta_j)(1-\eta_k+\eta_i-\eta_j) (1-\eta_k-\eta_i+\eta_j) }                \nonumber \\
\xi_{ij} &= &\frac{1}{4\sqrt{\eta_i\eta_j}}\sqrt{(1+\eta_k-\eta_i-\eta_j) (1+\eta_k+\eta_i+\eta_j)(1-\eta_k+\eta_i+\eta_j) (\eta_k+\eta_i+\eta_j-1) }             \nonumber \\
\end{eqnarray}
The six real parameters are $\eta_i $ and $\delta_i (i=1, 2, 3) $. In what follows we use 
$ (\rho+i\xi )_{ij}\equiv \eta_{ij} e^{i\phi_{ij}} $.
This unitary S-matrix has the following properties:  First, if any one of $\eta_i $ becomes unity, the other two become identical.  For example, if $\eta_3=1 $, then the requirement that all three off-diagonal $\eta_{i j} $ be $\geq 0 $ implies that $\eta_1=\eta_2\equiv\eta $. In this situation   channel 3 decouples from channels 1 and 2  and one  recovers the 2-channel form  (16)
for the S-matrix for these channels.  Second, if all $\eta_i\rightarrow 1 $, then all $\eta_{ij}\rightarrow 0 $ and the S-matrix becomes diagonal with diagonal entries $e^{2i\delta_i }$, i.e.,  each channel becomes elastic.
Now using the S-matrix of (20) in (13) we can write down the analogues of (17) and (18) as follows:
\begin{eqnarray}
 \sigma_1(s) =  \frac{1}{2}\eta_1 sin2\delta_1 M^0_1(s)   + \frac{1}{2}\eta_{12}sin(\phi_{12}+\delta_1+\delta_2) M^0_2(s)       \nonumber \\
+\frac{1}{2}\eta_{13}sin(\phi_{13}+\delta_1+\delta_3) M^0_3(s) \\
Im M_1(s) = \sigma_1(s) ~~~~~~~~~~~~~~~~~~~~~~~~~~~~~~~~~~~~~~~~~~~~~  \nonumber   \\
Re M_1(s) =  \frac{1}{2}(1+\eta_1 cos 2\delta_1) M_1^0(s) +\frac{1}{2}\eta_{12}cos (\phi_{12}+\delta_1+\delta_2) M^0_2(s)\\
 + \frac{1}{2}\eta_{13}cos (\phi_{13}+\delta_1+\delta_3) M^0_3(s). \nonumber
\end{eqnarray}
The analogous relations for other values of the index are written down by symmetry. Now, if the Pomeron were the sole contributor to the elastic amplitudes (the diagonal elements of the S-matrix), $\delta_i $ would be strictly zero and $\eta_i $ close to ( and less than ) unity. The fact that Regge exchanges contribute to the elastic amplitudes makes $\delta_i $ finite through small\cite{Zheng,NP}.  As $\sqrt{s} $ increases in the B mass region, the Regge-generated amplitudes move towards the base of the unitary circle  while the Pomeron
generated amplitudes do the same albeit at a much slower rate. Such a motion of the elastic amplitudes in the Argand diagram leads to an increase of $\eta_i $ towards unity and a decrease in the magnitudes of $\delta_i $.

\par
A statement as to what happens as $s\rightarrow \infty $ depends very much on the behavior of the Pomeron amplitude.  The Regge-exchange  generated  S-wave falls faster than  $s^{\alpha (0)-1} $.
If, as $s\rightarrow \infty $, the Pomeron "softens" such that $\alpha_P(0)\rightarrow 1 $, then the Pomeron-generated S-wave amplitude will also move along the imaginary axis of the Argand diagram 
towards  the base.  Thus as $ s\rightarrow\infty $,  all $\eta_i\rightarrow 1,  \eta_{ij}\rightarrow 0 $ and $\delta_i\rightarrow 0 $.   The scattering amplitude reaches the elastic, though a "no-scattering", limit. The S-matrix becomes a unit matrix and $Im M_i(s) \rightarrow 0, M_i(s)\rightarrow  M^{(0)}_i(s) $.  If, on the other hand, as $s\rightarrow \infty $ the Pomeron-generated S-wave amplitudes reach  fixed points on the imaginary axis, then all $\eta_i $ will approach values in the interval (0,1) with $2\delta_i=0 $ or $\pi $ depending on whether the fixed point is below or above the center of the unitary circle.  The S-matrix will approach a  real-symmetric form.  For three channels such an S-matrix has diagonal elements $ (\eta_1,  \eta_2,   \eta_3 ) $ and  off-diagonal elements $\eta_{ij} $
with
\begin{equation}
\sum_{i=1}^{3}\eta_i=1~~~~and~~~~\eta_{ij}=-\sqrt{(1-\eta_i) (1-\eta_j)}.
\end{equation}
This form can generalized to an $(n\times n) $ case by constraining $\eta_i$'s by requiring $\sum_{i=1}^{n}\eta_i = (n-2) $ and defining $\eta_{ij} $ as in (24).  For the $(3\times 3) $ case, if, say, $\eta_3=1 $, then $\eta_1=\eta_2=0 $ and channel 3 decouples from channels 1 and 2 which are now driven entirely by   inelastic channels.   If we assume that such a situation does not  arise then $\eta_i < 1$. This implies that elastic unitarity is not satisfied in any channel and inelastic channels must be there to implement unitarity.   Eq.(13) now implies that $\sigma_i(s)=0 ~ ( Im M_i(s) = 0 ) $ while (12) implies that each of $M_i(s) $ becomes a  real superposition of all $M^{(0)}_i(s) $.
\par
We conclude by a critique  of  the often-used adhoc procedure\cite{Zheng,NP,NRSX} to unitarize real amplitudes $\bf{M}^{(0)}(s) $ in presence of fsi given by (14).  For a  single channel, $S=exp(2i\delta) $, and hence $M(s)=exp(i\delta)M^0(s) $. This is said to be a statement of Watson's theorem. Eq.(14) is then a naive multi-channel  generalization of the single-channel case. Watson's theorem , however, only claims that in a single-channel case, the phase of the formfactor,  or decay amplitude,  be the same as the scattering phase. It is to be emphasized that the magnitude of the formfactor also changes as a result of fsi.  In general, for a single channel case the formfactors satisfy  Muskhelishvili-Omnes  equation\cite{M,O,B},
\begin{equation}
M(s) =  exp\{\frac{P}{\pi}\int \frac{\delta (s^{\prime})ds^{\prime} }{s^{\prime}-s}\} M^0(s)
exp(i\delta (s))
\end{equation}
where P signifies 'principal part'. Clearly, even for elastic scattering the magnitude of the formfactor changes -- a circumstance   assumption (14)  forbids.

In conclusion, we have investigated the issue of fsi in the limit of heavy quark masses using Pomeron and Regge exchanges  as guides to high energy scattering. In the B mass region  Pomeron exchange does not dominate as evidenced by the still-decreasing total cross-sections. 
As energy is increased ( but still in tens-of-GeV region), the S-matrix tends towards a real-symmetric form as the Regge-generated  partial-wave  amplitude drops with energy  faster than  $s^{\alpha_R(0)-1} $. Asymptotically, if $\alpha_P(0) $ remains at 1.08,  partial-wave unitarity is violated around $\sqrt{s}\sim 10^{11} GeV $. However, if the Pomeron were to "soften" such that $\alpha_P(0)\rightarrow 1 $ asymptotically then the S-matrix approaches a unit matrix--"no-scattering" limit. In this situation $M_i(s) $ become real with $ M_i(s)\rightarrow M^{(0)}_i(s) $.  If, on the other hand, the elastic amplitudes approach  fixed points  on the imaginary axis of the Argand diagram, the S-matrix becomes real-symmetric. Inelastic channels must be there to satisfy unitarity, yet they do not impart an imaginary part to the decay amplitude. Each $M_i(s) $ now becomes a real superposition of all $ M^{(0)}_i(s) $.
\par
This work was partially supported by an individual  research grant  from the Natural Sciences and Engineering Research Council of Canada  to ANK. 


\end{document}